\documentclass[pra,twocolumn,groupedaddress,showpacs,floatfix]{revtex4}
\usepackage{amsfonts}
\usepackage{amsmath}
\usepackage{graphicx}
\usepackage{mathrsfs}
\usepackage{bm}
\usepackage{amsbsy,amsmath}
\usepackage{color}
\usepackage{float}

\newcommand{\ket}[1]{|#1 \rangle}
\newcommand{\bra}[1]{\langle #1|}

\begin{document}

\title{An alternative power spectrum of the resonance fluorescence of atomic systems}
\author{Adam Stokes and Almut Beige}
\affiliation{The School of Physics and Astronomy, University of Leeds, Leeds LS2 9JT, United Kingdom}
\date{\today}

\begin{abstract}
We adopt an open quantum systems perspective to calculate the power spectrum associated with the electric field generated by an atomic dipole moment undergoing resonant laser-driving. This spectrum has a similar shape to the usual Mollow spectrum, but also has some distinct features. For sufficiently strong laser driving, both spectra have a symmetric triplet structure with a large central peak and two sidebands. However, the relative height of the sidebands to the central peak differs in each case. The two spectra also behave quite differently when the laser Rabi frequency is varied. Both spectra may be of interest in high-precision experiments into the quantum physics of atomic systems, especially artificial atoms. 
\end{abstract}

\maketitle

\section{Introduction}

Recent years have seen a wide range of high-precision experiments, which study the light-matter interactions of trapped ions \cite{RevMod}, single quantum dots \cite{Atature}, colour centres \cite{Jelezko}, and molecules on surfaces \cite{Hwang}. What all of these systems have in common is that an external laser excites a strongly-confined ground state electron into an excited state, which results in the spontaneous emission of photons. So-called artificial atoms, like quantum dots and colour centres, often have much stronger spontaneous decay rates $\Gamma$ and significantly smaller transition frequencies $\omega_0$ than real atoms and ions. As a result, they provide a new testing ground for standard quantum optical models. This includes models that take into account the decoherence of the atom-field system due to the presence of an external environment. An efficient way of obtaining insight into the structure and dynamics of laser-driven atomic systems in the presence of decohering environments, is to perform measurements on the electromagnetic signals associated with their resonance fluorescence. 

In this paper, we adopt an open systems perspective to investigate the power spectrum associated with the total electric field generated by a laser-driven atomic system capable of spontaneously emitting photons. As in classical physics, we define the power spectrum in terms of a correlation function $G(\tau)$. However, our quantum correlation function reflects a much richer inner dynamics of the atomic system. This is because quantum measurements strongly effect the state of the quantum system being measured. Measurement outcomes can therefore be highly correlated with previous measurements.

A noteworthy aspect of our approach is that the continuous loss of memory between the atom and the surrounding free radiation field is explicitly taken into account \cite{Kimble}. This is achieved by assuming the presence of a decoherence mechanism for the atom-field system. More specifically, we assume that the atom-field system is surrounded by a photon-absorbing environment. This photon-absorbing environment acts as a monitor of the radiation field, which at each $\Delta t$ time step performs a photon-number resolving measurement of the field \cite{Hegerfeldt93,Breuer}, and subsequently resets it into its vacuum state $\ket{0}$. We call the parameter $\Delta t$ that determines the frequency of these measurements the typical environmental response time \cite{Hegerfeldt93,Molmer,Carmichael,Zoller,Ourpaper}. Moreover, following Zurek \cite{Zurek2}, the field vacuum $\ket{0}$ can be termed the {\em einselected} state of the radiation field. In the absence of a photon source, but in the presence of a photon-absorbing environment, the vacuum is the only state that does not evolve in time. This makes it the preferred state into which the radiation field rapidly relaxes once a photon has been emitted by the atom.

As in our previous paper \cite{Ourpaper}, we avoid approximations whenever possible. For example, instead of simply applying the rotating wave approximation, we emphasize that it is possible to obtain an atom-field Hamiltonian without counter-rotating terms by means of a unitary transformation that (partially) diagonalises the Coulomb gauge Hamiltonian \cite{drummond,baxter,Ourpaper}. This is a crucial point within our model, because the effect of the photon-absorbing environment depends on the presence of counter-rotating terms in the interaction Hamiltonian. When such terms are present, the continuous resetting of the radiation field onto the vacuum state actually pumps free-energy into the atom-field system. This is because the presence of counter-rotating terms implies that the bare ground state $\ket{0,0}$ consisting of no photons and the atom in its ground state, has {\em greater} energy than the ground state of the interacting atom-field system.

When Markovian and rotating-wave approximations are avoided, the continuous resetting of the field onto the vacuum results in a build up of energy, which manifests itself in the form of a non-zero stationary state photon emission rate even in the absence of external driving \cite{andreas,Ourpaper}. Such emission rates would be visible in actual experiments \cite{Atature,Jelezko}. Thus, if we assume on physical grounds that an un-driven atom should not emit photons from its stationary state, then we must conclude that the photon-absorbing environment resets the field onto the vacuum state associated with the so-called rotating-wave Hamiltonian, which contains no counter-rotating terms \cite{Ourpaper}. 

The above assumption of continuous environmental resetting over a relatively short time scale, comprises an extremely strong physical constraint to be imposed on the atom-field system. We therefore expect it to have non-trivial consequences for the coarse-grained dynamics of observables pertaining to the radiation field. In this paper our aim is to investigate these consequences. We do this by calculating the power spectrum associated with the electric field generated by the laser-driven atomic source, and comparing this spectrum to the so-called Mollow spectrum \cite{Mollow}. Our hope is that our analysis provides new insights into the dynamics of laser driven atom-field systems in the presence of decohering environments. We find that the electric field spectrum as well as the Mollow spectrum have many similarities. For sufficiently strong laser driving both spectra exhibit a central peak with two sidebands, but the relative height of these peaks is different.

Before continuing we note that the electric field produced by a microscopic atomic source may be difficult to accurately measure in practice. Such a field is only likely to induce a very small electromotive force on a chosen test charge. The field is also expected to drop-off rapidly with increasing distance from the source. However, recent technological advances \cite{Schwab} have paved the way for observing extremely small forces. For example, Usenko {\em et al.}~\cite{Usenko} recently demonstrated the detection of sub-attonewton forces at milliKelvin temperatures by using a superconducting quantum interference device. Other authors employ quantum point contacts as sensitive displacement detectors in high precision experiments which aim at quantum limited displacement detection \cite{Field,Petta,Poggio2}. Moreover, the technology which is needed to combine a single quantum dot and an atomic force microscope cantilever in a single experimental setup has already become available \cite{Clerk}. 

There are five sections in this paper. In Section \ref{back}, we summarise the theoretical background of this paper, we provide a definition of the power spectrum of a classical signal, identify the electric field observable ${\bf E}({\bf x})$ which corresponds to the above mentioned rotating wave Hamiltonian, and discuss the validity of the two-level approximations and the use of a master equation. In Section \ref{field}, we derive the power spectrum associated with the electric field observable ${\bf E}({\bf x})$ of a laser-driven atomic two-level system with spontaneous photon emission and show that this spectrum has a central peak at the atomic transition frequency $\omega_0$. Section \ref{Mollow} compares this spectrum with the usual Mollow spectrum \cite{Mollow}. Finally, we summarise our findings in Section \ref{conc}. 

\section{Theoretical background} \label{back}

\subsection{The classical correlation function and spectrum}

In classical physics, the power spectrum $S(\omega )$ of a signal $f(t)$ equals the modulus squared of its Fourier transform $\tilde f(\omega)$. To prevent $S(\omega)$ from tending to infinity for a wide range of signals, one defines
\begin{eqnarray} \label{defa}
S (\omega) &\equiv & | \tilde f(\omega) |^2  
\end{eqnarray}
with the truncated and time averaged Fourier transform $\tilde f(\omega)$ given by
\begin{eqnarray} \label{defb}
\tilde f(\omega) &=& \lim_{T \to \infty} {1 \over \sqrt{2T}} \int_{-T}^T {\rm d}t \,  {\rm e}^{-{\rm i} \omega t} f(t) \, . 
\end{eqnarray}
For a real signal $f(t)$, this definition implies
\begin{eqnarray}
S (\omega) = \lim_{T \to \infty} {1 \over 2T} \int_{-T}^T {\rm d} t \int_{-T}^T {\rm d} t' \, {\rm e}^{-{\rm i} \omega (t - t')} f(t) f(t') \, , 
\end{eqnarray}
where the limits on the right hand side are assumed to exist. After making the substitution $\tau = t - t'$, we find that the power spectrum of a signal equals the Fourier transform of its two-time correlation function $G(\tau)$,
\begin{eqnarray} \label{spectrum}
S (\omega) &=& \int_{- \infty}^\infty {\rm d} \tau \, {\rm e}^{-{\rm i} \omega \tau} \, G(\tau) 
\end{eqnarray}
with $G(\tau)$ defined by
\begin{eqnarray} \label{Gtau}
G (\tau) &=& \lim_{T \to \infty} {1 \over 2T} \int_{-T}^T {\rm d}t \, f(t) \, f(t+\tau) \, .
\end{eqnarray}
We are interested in the total electric field generated by the atom, so for a detector located at position ${\bf x}$, we consider the signal 
\begin{align}
{\bf f}_E(t,{\bf x}) =& \langle {\bf E} ({\bf x}) \rangle_t \, .
\end{align}
Our hope is that an analysis of the spectrum associated with ${\bf f}_E(t,{\bf x})$ might provide new insights into the dynamics of atom-field systems in the presence of decohering environments.

\subsection{The signal in the presence of a photon-absorbing environment}

We assume in the following that a wire at a position ${\bf x}$ and time $t$, performs a direct measurement of the electric field generated by the atomic system. In this section we identify the {\em atomic} operator that represents this electric field, under the assumption of continuous environmental resetting of the radiation field onto its vacuum state. To begin with, we consider an atomic dipole with canonical operators ${\bf r}$ and ${\bf p}$ satisfying
\begin{align}
[r_i,p_j]=i\hbar\delta_{ij} \, .
\end{align}
The dipole interacts with a transverse electromagnetic field with canonical field operators ${\bf A}_{\rm T}$ and ${\bm \Pi}_{\rm T}$ satisfying
\begin{align}
[{\rm A}_i({\bf x}),\Pi_j({\bf x}')] = {\rm i} \hbar \, \delta_{\it ij}^{\rm T}({\bf{x}}-{\bf{x}}') \, ,
\end{align}
where $\delta^{\rm T}$ denotes the transverse delta function. These fields support the following mode expansions
\begin{align}\label{modeex}
{\bf A}({\bf x}) =&  \int d^3k \sum_{\lambda} \sqrt{ \frac{\hbar}{2\epsilon_0\omega_k (2\pi)^3}}  \, {\rm{\bf e}}_{{\bf k}\lambda } \, a_\lambda ({\bf k})  \, {\rm e}^{{\rm i} {\bf k} \cdot {\bf x}}+ {\rm H.c.} \, , \nonumber \\
{\bf \Pi}({\bf x}) =& - {\rm i}  \int d^3k \sum_{\lambda} \sqrt{ \frac{\hbar \epsilon_0 \omega_k }{2 (2\pi)^3}}{\rm {\bf e}}_{{\bf k}\lambda } \, a_\lambda ({\bf k})  \, {\rm e}^{{\rm i}{\bf k}\cdot{\bf x}} + {\rm H.c.} \, , 
\end{align}
where the $a_\lambda ({\bf k})$ and $a_\lambda^\dagger({\bf k})$ are photon annihilation and creation operators satisfying the bosonic commutation relation
\begin{align}\label{craadagger}
[a_\lambda ({\bf k}) ,a_{\lambda'}^\dagger ({\bf k}')] = \delta_{\lambda\lambda'}\delta({\bf k}-{\bf k}') \, .
\end{align}
Each vector ${\bf e}_{{\bf k}\lambda}$ with $\lambda=1,2$  in Eq.~(\ref{modeex}) is a unit vector orthogonal to ${\bf k}$. Moreover, $\omega_k \equiv c|{\bf k}|$.

Let us consider the negative transverse displacement field defined by $-{\bf D}_{\rm T} = -\epsilon_0{\bf E}_{\rm T}-{\bf P}_{\rm T}$. In the Coulomb gauge the canonical momentum ${\bm \Pi}_{\rm T}$ represents the negative of the transverse electric field $-\epsilon_0{\bf E}_{\rm T}$. The field ${\bf P}_{\rm T}$ meanwhile is the transverse multipolar polarisation field defined by
\begin{align}
P_{{\rm T},i}({\bf x}) \equiv -e\int_0^1 d\lambda \, r_j \delta_{ij}^{\rm T}({\bf x}-\lambda{\bf r}) \, .
\end{align}
For a neutral system of charges such as the atomic system we are considering, the displacement field is entirely transverse; ${\bf D}={\bf D}_{\rm T}$. In addition, in the electric dipole approximation (EDA), which we will employ throughout this paper, the atomic system is taken as coupling to the field at the atomic centre-of-mass position, which we can take as the origin with coordinates ${\bf 0}$. In the EDA, the electric polarisation field ${\bf P}({\bf x})$ associated with the atomic dipole $-e{\bf r}$ is given by $-e{\bf r} \, \delta({\bf x})$, which is clearly localised at the origin. As such, for ${\bf x}\neq {\bf 0}$ we have in the EDA that 
\begin{align}
\epsilon_0{\bf E}_{\rm T}({\bf x})+{\bf P}_{\rm T}({\bf x}) \equiv {\bf D}_{\rm T}({\bf x}) \equiv {\bf D}({\bf x}) \equiv \epsilon_0{\bf E}({\bf x}).
\end{align}
Next we determine the appropriate operator with which we must represent this observable.

Here we are assuming that the radiation field is continuously reset onto the field vacuum associated with the rotating-wave Hamiltonian. In order to determine the effect this assumption has on the dynamics of the electric field, we must identify within the rotating-wave representation, the operator that represents the physical field $\epsilon_0{\bf E}_{\rm T}+{\bf P}_{\rm T}$. The rotating-wave representation is related to the Coulomb gauge representation via a unitary transformation $R$ which in the EDA is given by \cite{Ourpaper}
\begin{eqnarray}\label{ralphak}
R_{\{\alpha_k\}} \equiv \exp \left({{\rm i} e \over \hbar} {\bf A}_{\alpha_k}({\bf 0}) \cdot {\bf r} \right)
\end{eqnarray}
where
\begin{align} \label{add2}
{\bf A}_{\alpha_k}({\bf 0}) \equiv \int d^3k \sum_{\lambda} \sqrt{ \frac{\hbar}{2\epsilon_0\omega_k(2\pi)^3}}  \, \alpha_k \, {\rm{\bf e}}_{{\bf k}\lambda } \, a_{{\bf k}\lambda } + {\rm H.c.} 
\end{align}
and
\begin{eqnarray} \label{alphak}
\alpha_k \equiv {\omega_0 \over \omega_0 + \omega_k} \, .
\end{eqnarray}
The quantity $\hbar\omega_0$ is a positive constant that will be identified as the energy between the ground and excited states of the atom, once the two-level approximation has been made. If we denote the Coulomb gauge Hamiltonian by $H$, then within the two-level approximation and with the above choice for the parameter $\alpha_k$, the Hamiltonian 
\begin{eqnarray} 
H_{\rm rot} \equiv R_{\{\alpha_k\}} H R_{\{\alpha_k\}}^{-1} 
\end{eqnarray} 
possesses no counter-rotating terms in its interaction component.

In order to identify the operator representing the total electric field in the rotating-wave representation we note that in the Coulomb gauge $-\epsilon_0{\bf E}_{\rm T} ={\bm \Pi}_{\rm T}$ and in the EDA
\begin{align}
{\bf P}_{\rm T}({\bf x}) = -{e\over (2\pi)^3} \int d^3k \sum_{\lambda}{\bf e}_\lambda ({\bf k})({\bf e}_\lambda ({\bf k})\cdot {\bf r})e^{i{\bf k}\cdot {\bf x}}.
\end{align}
Thus, noting that $R_{\{\alpha_k\}}$ commutes with ${\bf P}_{\rm T}$, the total electric field at ${\bf x}\neq {\bf 0}$ is given in the rotating-wave representation by
\begin{align}\label{erot}
\epsilon_0{\bf E}({\bf x}) =& -R{\bm \Pi}_{\rm T}({\bf x})R^{-1} +{\bf P}_{\rm T}({\bf x}) \, .
\end{align}
Hence, for ${\bf x}\neq {\bf 0}$
\begin{align}\label{erotb}
\epsilon_0{\bf E}({\bf x}) =& -{\bm \Pi}_{\rm T}({\bf x}) -{e\over (2\pi)^3} \int d^3k \nonumber \\ 
& \times \sum_{\lambda}(1-\alpha_k){\bf e}_\lambda ({\bf k})({\bf e}_\lambda ({\bf k})\cdot {\bf r})e^{i{\bf k}\cdot {\bf x}} \, .
\end{align}

The photon-absorbing environment that we assume is present acts as a monitor, which at each $\Delta t$ time step performs a photon number measurement on the radiation field \cite{Hegerfeldt93,Breuer}. For sufficiently small $\Delta t$ there can be at most one-photon within the radiation field at any given instant $t=n\Delta t$ \cite{Hegerfeldt93,Molmer,Carmichael,Zoller,Ourpaper}. The state after an environmental measurement is therefore either the vacuum or a one-photon state. In the latter case the environment immediatey resets the field into the vacuum. This model of environmental decoherence allows us to assume for the purposes of calculating expectation values, that the density matrix of the atom-field system in the Schr\"odinger picture at time $t=n \Delta t$, is given by 
\begin{align}
\rho(t) = \rho_{\rm A}(t)\otimes \rho_{\rm F}(t)
\end{align}
where $\rho_{\rm F}(t)$ denotes a (suitably normalised) classical mixture of the vacuum and one-photon states;
\begin{align}
\rho_{\rm F}(t) = p_0(t)\ket{0}\bra{0} + \sum_{{\bf k}\lambda} p_{{\bf k}\lambda}(t) \ket{1_{{\bf k}\lambda} }\bra {1_{{\bf k}\lambda} }.
\end{align}
Since with $\rho_{\rm F}(t)$ defined above ${\rm Tr}_{\rm F}(\rho_{\rm F}(t){\bm \Pi}_{\rm T}({\bf x}))=0$, the only nonzero contribution from ${\bf E}({\bf x})$ in Eq.~(\ref{erot}) to the signal ${\bf f}_E(t,{\bf x}) = \langle {\bf E({\bf x})} \rangle_t$ comes from the generalised polarisation term
\begin{align}\label{erot2}
w_{ij}({\bf x})r_j\equiv {1\over (2\pi)^3} \int d^3k \sum_{\lambda}(1-\alpha_k)e^i_\lambda ({\bf k})e^j_\lambda ({\bf k}) r_je^{i{\bf k}\cdot {\bf x}}
\end{align}
in which the repeated roman index is to be summed, and where the $w_{ij}$ are functions of the classical variable ${\bf x}$ denoting the detector position. The functions $w_{ij}$ can be reduced by performing the angular integration and polarisation summation in Eq.~(\ref{erot2}), which gives
\begin{align}\label{erot2b}
w_{ij}({\bf x}) = {1\over 2\pi^2} (-\delta_{ij} \partial^2 + \partial_i\partial_j )\int_0^\infty d\omega_k \, {\sin(\omega_k |{\bf x}| /c) \over (\omega_0 + \omega_k)|{\bf x}|}.
\end{align}
For $\omega_0\neq 0$ the integral over frequency can only be expressed in terms of special functions. However, setting $\omega_0 \equiv 0$ the frequency integration in Eq. (\ref{erot2b}) evaluates to $\pi/ 2|{\bf x}|$. Eq. (\ref{erot2}) then gives $-1/e$ times $P_{{\rm T},i}^{\rm EDA}({\bf x})$, which is the usual multipolar polarisation field in the EDA. Performing the differentiations one then obtains a sum of three terms which vary as $|{\bf x}|^{-1}, |{\bf x}|^{-2}$ and $|{\bf x}|^{-3}$ respectively \cite{craig}. A similar $|{\bf x}|$ dependence is expected when $\omega_0 \neq 0$.

The $w_{ij}$ are c-number functions. On the other hand ${\bf r}$ denotes the atomic operator whose expectation value must be calculated in order to evaluate the signal 
\begin{align}\label{sig1}
\langle E_i({\bf x})\rangle_t = -e w_{ij}({\bf x})\langle r_j \rangle_t\, .
\end{align}
This will be done in what follows using the two-level approximation and the standard Born-Markov quantum optical master equation.

\subsection{Two-level approximation and the effective signal}

We restrict our attention now to two atomic levels $\ket{0}$ and $\ket{1}$. The operators $\sigma^+ = \ket{1}\bra{0}$ and $\sigma^- = \ket{0}\bra{1}$ raise and lower these atomic levels respectively. The atomic dipole moment $-e{\bf r}$ can be written as
\begin{align}
-e{\bf r} = {\bf d}\sigma_{\rm x},~~~~~~{\bf d} = -e\bra{0}{\bf r}\ket{1}
\end{align}
where we have assumed for simplicity that ${\bf d}$ is real, and $\sigma_{\rm x} \equiv \sigma^+ + \sigma^-$. Within the two-level approximation we find that the Schr\"odinger picture signal in Eq.~(\ref{sig1}) becomes 
\begin{eqnarray} \label{f(t)a}
\langle E_i({\bf x})\rangle_t = w_{ij}({\bf x}) d_j \langle \sigma_{\rm x} \rangle_t  
\end{eqnarray}
Thus, up to the additional factors $w_{ij}({\bf x})d_j$, which do not depend on the atomic dynamics, the signal we are interested in is
\begin{eqnarray} \label{f(t)}
f_E(t) = \langle \sigma_{\rm x} \rangle_t.
\end{eqnarray}
As we shall see below, the power spectrum $S(\omega)$ associated with the signal in Eq.~(\ref{f(t)}) is a direct measure for the coherence of the atomic source and not a measure of its intensity. 

\subsection{Atomic master equations} \label{meq}

With the aim of calculating $\langle \sigma_{\rm x} \rangle_t$, we solve in the following the standard Born-Markov quantum optical master equation of a resonantly driven atomic two-level system with spontaneous decay rate $\Gamma$;
\begin{eqnarray}\label{rhodotI2}
\dot \rho_{\rm A}(t) &=& - {{\rm i} \over \hbar} \, [ \hbar \omega_0 \, \sigma^+ \sigma^- + H_{\rm L}(t) , \rho_{\rm A} (t) ] \nonumber \\
&& + \Gamma \, \sigma^- \, \rho_{\rm A}(t) \, \sigma^+ - {1 \over 2} \Gamma \, \left\{ \sigma^+ \sigma^- , \rho_{\rm A}(t) \right \} \, , \nonumber \\
 H_{\rm L}(t) &=& {1 \over 2} \hbar \Omega \, \sigma^+  \, {\rm e}^{{\rm i}\omega_0 t} + {\rm H.c.} 
\end{eqnarray}
Here $\hbar \omega_0$ denotes the energy difference between the atomic levels, and $\Omega$ is a real laser Rabi frequency. Moving into the interaction picture with respect to 
\begin{eqnarray} \label{H0}
H_0 &=& \hbar \omega_0 \, \sigma^+ \sigma^- \, , 
\end{eqnarray}
the above master equation becomes time-independent and can be solved analytically. We denote the stationary solution $\rho_{\rm ss}$, and use the notation $\rho_{ij} \equiv \langle i| \rho_{\rm AI} |j \rangle$ for the elements of the interaction picture density matrix $\rho_{\rm AI}$. Due to the relations $\rho_{10} = \rho_{01}^*$ and $\rho_{11} = 1 - \rho_{00}$ with $\rho_{00}$ real, the master equation has only three independent real solutions.

We proceed now in solving the interaction picture atomic dynamics relative to the stationary state. This gives
\begin{eqnarray} \label{rho22}
{\rm Re} \, \rho_{01} (t+\tau) &=& {\rm e}^{- \Gamma \tau/2} \, {\rm Re} \, \rho_{01} (t), 
\end{eqnarray}
and
\begin{align} \label{rho22x}
\left(  \begin{array}{c} \rho_{00} (t+\tau) \\ {\rm Im} \, \rho_{01} (t+\tau) \end{array}  \right) 
=& \, A(\tau)\left( \begin{array}{c} \rho_{00}(t) - \rho_{00}^{\rm ss} \\ {\rm Im} \, \rho_{01}(t) - {\rm Im} \, \rho_{01}^{\rm ss} \end{array} \right) \nonumber \\ &+  \left( \begin{array}{c} \rho_{00}^{\rm ss} \\ {\rm Im} \, \rho_{01}^{\rm ss} \end{array} \right)\,
\end{align}
where
\begin{align}\label{A}
A(\tau)\equiv\left[ I\cos\mu \tau - {1 \over 4 \mu} \left( \begin{array}{cc} \Gamma & 4 \Omega \\ - 4 \Omega & - \Gamma \end{array} \right) \, \sin\mu \tau \right] {\rm e}^{- 3 \Gamma \tau/4}
\end{align}
in which $I$ is the $2\times 2$ identity matrix, and
\begin{eqnarray} \label{mu}
\mu &\equiv & {1 \over 4} \sqrt{16 \Omega^2 - \Gamma^2}.
\end{eqnarray}

The stationary state matrix elements appearing in Eq. (\ref{rho22x}) are given by
\begin{align}\label{st} 
\rho_{00}^{\rm ss} =  {\Gamma^2 + \Omega^2 \over \Gamma^2 + 2 \Omega^2}, ~~
{\rm Im} \, \rho_{01}^{\rm ss} = {\Gamma \Omega \over \Gamma^2 + 2 \Omega^2},~~ {\rm Re} \, \rho_{01}^{\rm ss} = 0.
\end{align}
The atomic system under consideration possesses a stationary state only in the interaction picture. In the Schr\"odinger picture, the off-diagonal matrix elements of this state become time-dependent. As a result we choose to remain in the interaction picture wherein the observable of interest $\sigma_{\rm x}$ becomes time-dependent. 

\section{The power spectrum of the electric field} \label{field}

\subsection{Relevant interaction picture observables and states}

In this section we look more closely at the power spectrum associated with the signal $f_E(t)$ in Eq.~(\ref{f(t)}). To calculate $G(\tau)$ we exploit the fact that the system possesses a stationary state in the interaction picture. Within the interaction picture the operator $\sigma_{\rm x}$ becomes
\begin{eqnarray}
\sigma_{\rm x} (t) &=& \sigma^+ \, {\rm e}^{-{\rm i} \omega_0 t} + \sigma^- \, {\rm e}^{{\rm i} \omega_0 t}\, .
\end{eqnarray}
The eigenvectors of this operator are
\begin{eqnarray} \label{xx}
|\lambda_{0,1} (t) \rangle &=& {1 \over \sqrt{2}}\left( |0 \rangle \pm {\rm e}^{{\rm i} \omega_0 t} \, |1 \rangle \right) \, ,
\end{eqnarray}
which clearly oscillate in time. We denote the projection operator onto the time-dependent state $|\lambda_i(t) \rangle$
\begin{eqnarray}
I \!\! P_i (t) &=& |\lambda_i(t) \rangle \langle \lambda_i(t) | \, .
\end{eqnarray}
The eigenvalues of $\sigma_{\rm x}$ remain unchanged by the (unitary) transformation defining the interaction picture, so the eigenvalues of $\sigma_{\rm x}(t)$ are simply
\begin{eqnarray} \label{yy}
\lambda_{0,1} &=& \pm 1 \, . 
\end{eqnarray}
In terms of the $\lambda_i$ and $I \!\! P_i (t)$, the operator $\sigma_{\rm x}(t)$ affords the spectral representation
\begin{align}\label{sigspec}
\sigma_{\rm x}(t) = \sum_{i=0,1} \lambda_i I \!\! P_i (t).
\end{align}

\subsection{The quantum correlation function}

In order to calculate the correlation function $G(\tau)$ in Eq.~(\ref{Gtau}) we wish to interpret it in a way that is consistent with the probabilistic nature of quantum theory rather than classical theory. The product $f(t)f(t+\tau)$ in Eq. (\ref{Gtau}) corresponds to two consecutive measurements --- one made at time $t$, and then another made at time $t+\tau$. If the measured signal $f(t)$ were classical, then the measurement made at $t$ would not alter the state of the atom. The same state would then be propagated up to time $t+\tau$ to determine the signal $f(t+\tau)$. However, in quantum theory the state immediately after the measurement at $t$ is a collapsed state that depends on the outcome of the measurement made at $t$. As a result, the quantity $f(t+\tau)f(t)$ behaves quite differently in the quantum setting.

The signal in which we are interested is $f_E(t) = \langle\sigma_{\rm x}\rangle_t$ given in Eq. (\ref{f(t)}). In the following we envisage a specific operational procedure, whose aim is to determe the quantum correlations in the signal $f_E(t)$ between two different times. With regard to this procedure the product $f_E(t)f_E(t+\tau)$ is {\em not} simply interpreted as the product $\langle\sigma_{\rm x}\rangle_t\langle\sigma_{\rm x}\rangle_{t+\tau}$. We use the notation $\langle {\mathcal O};\rho \rangle$ to denote the expectation value of the operator ${\mathcal O}$ taken in the state $\rho$.

Let us consider first measurements of $\sigma_{\rm x}(t)$ made at time $t$ on an ensemble of identical atomic systems described by the density matrix $\rho_{\rm AI}(t)$. The average value of the observable $\sigma_{\rm x}(t)$ obtained via repeated measurements over the entire ensemble would be
\begin{align}\label{sigxt}
\langle\sigma_{\rm x}(t); \rho_{\rm AI}(t) \rangle_t = \sum_{i=0,1} \bra{\lambda_i(t)} \rho_{\rm AI}(t) \ket{\lambda_i(t)}\lambda_i
\end{align}
where Eq. (\ref{sigspec}) has been used. The summand in the above expression is the product of two numbers. The first
\begin{align}\label{rhoi}
\rho_{{\rm AI},i}(t) \equiv \bra{\lambda_i(t)} \rho_{\rm AI}(t) \ket{\lambda_i(t)}
\end{align}
represents the relative size of the subensemble of systems for which $\lambda_i$ was the measurement's outcome. The outcome $\lambda_i$, satisfies
\begin{align}\label{lami}
\lambda_i = \langle \sigma_{\rm x}(t); I \!\! P_i (t) \rangle_t\, ,
\end{align}
which simply states that the average value of measurements made at time $t$, taken over the subensemble for which $\lambda_i$ was the outcome measured, is $\lambda_i$ itself.

Let us now consider measurements made at time $t+\tau$. The subensemble of systems for which the outcome $\lambda_i$ was obtained for the measurements made at $t$, is described by the density matrix $\rho_{\rm AI}(t) =I \!\! P_i (t)$. The same subensemble at time $t+\tau$ is therefore described by
\begin{eqnarray}\label{exp1}
\rho_{\rm A I}(t+\tau)= {\cal T}_\tau (I \!\! P_i (t)) \,
\end{eqnarray}
where the superoperator ${\cal T}_\tau$ summarises the atomic time evolution in Eqs.~(\ref{rho22}) and (\ref{rho22x}). For this subensemble the average value of the measurements made at time $t+\tau$ is
\begin{align}\label{exp2}
\langle \sigma_{\rm x}(t+\tau) ; {\cal T}_\tau (I \!\! P_i (t)) \rangle_{t+\tau} &= \nonumber \\ &  \hspace*{-3mm} \sum_{j=0,1} \lambda_j {\rm Tr}\left[ I \!\! P_j (t+\tau)  {\cal T}_\tau (I \!\! P_i (t))\right],
\end{align}
Using Eqs.~(\ref{sigxt}) and (\ref{lami}), the product of the average value measured at time $t$ with that measured at time $t+\tau$, taken over the subensemble of systems for which $\lambda_i$ was the outcome measured at time $t$, is the product
\begin{align}\label{Gi}
G_i(t,\tau) \equiv \langle \sigma_{\rm x}(t); I \!\! P_i (t) \rangle_t \langle \sigma_{\rm x}(t+\tau) ; {\cal T}_\tau (I \!\! P_i (t)) \rangle_{t+\tau}\, .
\end{align}
In the following, we interpret the product $f_E(t+\tau)f_E(t)$ as a sum of the $G_i$ in Eq.~(\ref{Gi}), with each $G_i$ weighted according to the relative size of the subensemble for which the outcome $\lambda_i$ was found in the measurement at time $t$. The appropriate weighting factors are nothing but the $\rho_{{\rm AI},i}(t)$ defined in Eq.~(\ref{rhoi}). More succinctly, we assume that
\begin{align}\label{wsum}
f_E(t+\tau)f_E(t) &= \nonumber \\ & \hspace*{-2cm} \sum_{i=0,1}\rho_{{\rm AI},i}(t) \langle \sigma_{\rm x}(t); I \!\! P_i (t) \rangle_t \langle \sigma_{\rm x}(t+\tau) ; {\cal T}_\tau (I \!\! P_i (t)) \rangle_{t+\tau}\, .
\end{align}
The above expression provides a measure of correlations between measurements made at two different times $t$ and $t+\tau$. It is based on nothing but {\em conditional expectation values}, with the averages calculated at time $t+\tau$ conditioned upon the outcomes of the measurements made at the earlier time $t$. This constitutes one possible quantum mechanical extension of the classical correlation function $f_E(t+\tau)f_E(t)$.

Using Eqs.~(\ref{rhoi}), (\ref{lami}) and (\ref{exp2}) we can now write Eq.~(\ref{wsum}) as
\begin{align}\label{ff}
f_E(t+\tau)f_E(t) &= \nonumber \\ & \hspace*{-1.2cm}
\sum_{i,j=0,1} \lambda_i \lambda_j {\rm Tr} \left[ \, I \!\! P_j (t+\tau) \, {\cal T}_\tau \left( I \!\! P_i (t) \rho_{\rm AI}(t) I \!\! P_i (t) \right) \, \right].
\end{align}
The correlation function we are interested in is defined classically in Eq.~(\ref{Gtau}), but it can now be given in the quantum setting using Eq.~(\ref{ff}) as
\begin{eqnarray} \label{G5}
G(\tau) &=& \lim_{T \to \infty} {1 \over 2T} \, \int_{-T}^T {\rm d} t \, \sum_{i,j=0,1} \lambda_i \lambda_j \nonumber \\
&& \times {\rm Tr} \left[ \, I \!\! P_j (t+\tau) \, {\cal T}_\tau \left( I \!\! P_i (t) \rho_{\rm AI}(t) I \!\! P_i (t) \right) \, \right]\, . ~~
\end{eqnarray}
Now we have all the equations we need to calculate the power spectrum of the electric field.

\subsection{The power spectrum of the electric field} \label{main}

As we have seen in Section \ref{back}, the density matrix of the atom rapidly reaches a stationary state $\rho_{\rm ss}$. The time averaging $T\to \infty$ in Eq.~(\ref{G5}) implies that the general state $\rho_{\rm AI}(t)$ with which the quantity in Eq.~(\ref{G5}) is to be calculated, can be taken as $\rho_{\rm ss}$, since this is the state of the atomic system at almost all tmes. Evaluating the integrand in Eq.~(\ref{G5}) with the help of Eqs.~(\ref{rho22})--(\ref{st}), yields
\begin{align}\label{blah}
&G(\tau)= \lim_{T \to \infty} {1 \over 2T} \, \int_{-T}^T {\rm d} t \, \big[ \cos\omega_0 t\cos\omega_0(t+\tau) e^{-\Gamma\tau/2} \nonumber \\ &+ \sin\omega_0 t\sin\omega_0(t+\tau) \big( A_{11}(\tau)+2{\rm Im}\, \rho_{01}^{\rm ss} \nonumber \\ &\times \left\{A_{10}(\tau)\left[1-2\rho_{00}^{\rm ss}\right] +2{\rm Im}\, \rho_{01}^{\rm ss}\left[1-A_{11}(\tau)\right]\right\}\big)\big].
\end{align}
It is straightforward to carry out the time average in Eq. (\ref{blah}), which using the equalities
\begin{align}
\cos\omega_0 \tau &= \lim_{T\to \infty} {1\over T} \int_{-T}^{T} dt \, \cos\omega_0 t\cos\omega_0(t+\tau) \nonumber \\ &=  \lim_{T\to \infty} {1\over T}\int_{-T}^{T} dt\, \sin\omega_0 t\sin\omega_0(t+\tau)
\end{align}
gives
\begin{align}
G(\tau) =& {1\over 2}\cos\omega_0\tau\big[e^{-\Gamma\tau/2} + A_{11}(\tau) + 2{\rm Im}\, \rho_{01}^{\rm ss} \nonumber \\ & \hspace*{-0.4cm} \times \left(A_{10}(\tau)[1-2\rho_{00}^{\rm ss}]+2{\rm Im}\, \rho_{01}^{\rm ss} [1-A_{11}(\tau)] \right)\big].
\end{align}
Finally, using Eqs. (\ref{A}) and (\ref{st}) the above expression can be written
\begin{align}\label{G6}
G(\tau) &= {1 \over 2}  \left[ \, {\rm e}^{- \Gamma \tau/2} + \left( \beta^+ \, {\rm e}^{ {\rm i} \mu\tau} + \beta^- \, {\rm e}^{- {\rm i} \mu\tau} \right) {\rm e}^{- 3 \Gamma \tau /4} \right. \nonumber \\ 
& \hspace*{0.5cm} \left. + {4\Gamma^2 \Omega^2 \over (\Gamma^2 + 2 \Omega^2)^2} \, \right] \, \cos \omega_0 \tau 
\end{align}
where $\mu$ is defined in Eq.~(\ref{mu}), and the coefficients $\beta^\pm$ are given by
\begin{eqnarray}
\beta^\pm &\equiv& {\Gamma^4 + 4 \Omega^4 \over 2 (\Gamma^2 + 2 \Omega^2)^2} \mp {{\rm i} \Gamma \over 8 \mu} \left[ 1 - {12 \Gamma^2 \Omega^2 \over (\Gamma^2 + 2 \Omega^2)^2} \right] \, . ~~~~~
\end{eqnarray}
Substituting this result into Eq.~(\ref{spectrum}) and neglecting a sharp $\delta$-peak due to the last term in Eq.~(\ref{G6}), we finally obtain the power spectrum $S(\omega)$ of the electric field generated by a resonantly-driven atomic system,
\begin{eqnarray}\label{fl1yx} 
S(\omega) &=& {2\Gamma \over \Gamma^2 + 4 \delta^2} + 2 {\rm Re} \left( { 3\Gamma - 4 {\rm i}(\delta + \mu) \over 9\Gamma^2 + 16 (\delta +\mu)^2} \, \beta^+ \right) \nonumber \\
&& + 2 {\rm Re} \left( {3\Gamma - 4 {\rm i}(\delta - \mu) \over 9\Gamma^2 + 16 (\delta -\mu)^2} \, \beta^- \right) \, ,
\end{eqnarray}
where
\begin{eqnarray} \label{delta}
\delta \equiv  \omega - \omega_0 \, .
\end{eqnarray}
As we shall see in the next section, this spectrum is different to the Mollow spectrum.

\section{Comparison with the Mollow spectrum} \label{Mollow}

\subsection{The Mollow spectrum}

As an alternative to the power spectrum considered here, laser-driven atomic two-level systems are often characterised by the so-called Mollow triplet of resonance fluorescence \cite{cohen1,Breuer}. Since the spectrum we consider has many similarities with Mollow's spectrum, we briefly summarise the main characteristics of the latter. It is defined as the Fourier transform of the stationary state correlation function
\begin{eqnarray} \label{GMol}
G_{\rm Mol}(\tau) &=& \langle \sigma^+(t+\tau)\sigma^-(t)\rangle_{\rm ss} \nonumber \\
&=& \langle \sigma^+(\tau)\sigma^-(0)\rangle_{\rm ss}
\end{eqnarray}
with the last equality following from the homogeneity in time of stationary correlation functions. Calculating the expectation value in Eq.~(\ref{GMol}) with the help of the master equations presented in the previous section yields \cite{Breuer}
\begin{eqnarray}\label{fl2} 
S_{\rm Mol}(\omega) &=& {2 \Omega^2 \over \Gamma^2 +2 {\Omega}^2} \left[ {4 \Gamma \over \Gamma^2 + 4 \delta^2} \right. \nonumber \\
&& + 4 {\rm Re} \left( { 3\Gamma - 4 {\rm i}(\delta + \mu) \over 9\Gamma^2 + 16 (\delta +\mu)^2} \, \beta_{\rm Mol}^+ \right) \nonumber \\
&& \left. + 4 {\rm Re} \left( {3\Gamma - 4 {\rm i}(\delta - \mu) \over 9\Gamma^2 + 16 (\delta -\mu)^2} \, \beta_{\rm Mol}^- \right) \right] ~~~
\end{eqnarray}
with the frequencies $\mu$ and $\delta$ defined in Eqs.~(\ref{mu}) and (\ref{delta}) respectively, and with
\begin{eqnarray}\label{p2} 
\beta_{\rm Mol}^\pm &\equiv & - {\Gamma^2 - 2 \Omega^2 \over 4(\Gamma^2 +2\Omega^2)} \mp {{\rm i} \Gamma\over 16 \mu} \left[ 1 -  {12 \Omega^2 \over \Gamma^2 +2\Omega^2} \right] \, . ~~~~
\end{eqnarray}
This spectrum has its maximum at the atomic transition frequency $\omega_0$. 

\subsection{Comparison of both spectra}

\begin{figure}[t]
\begin{minipage}{\columnwidth}
\begin{center}
\hspace*{-1.2cm}
\includegraphics[scale=.8]{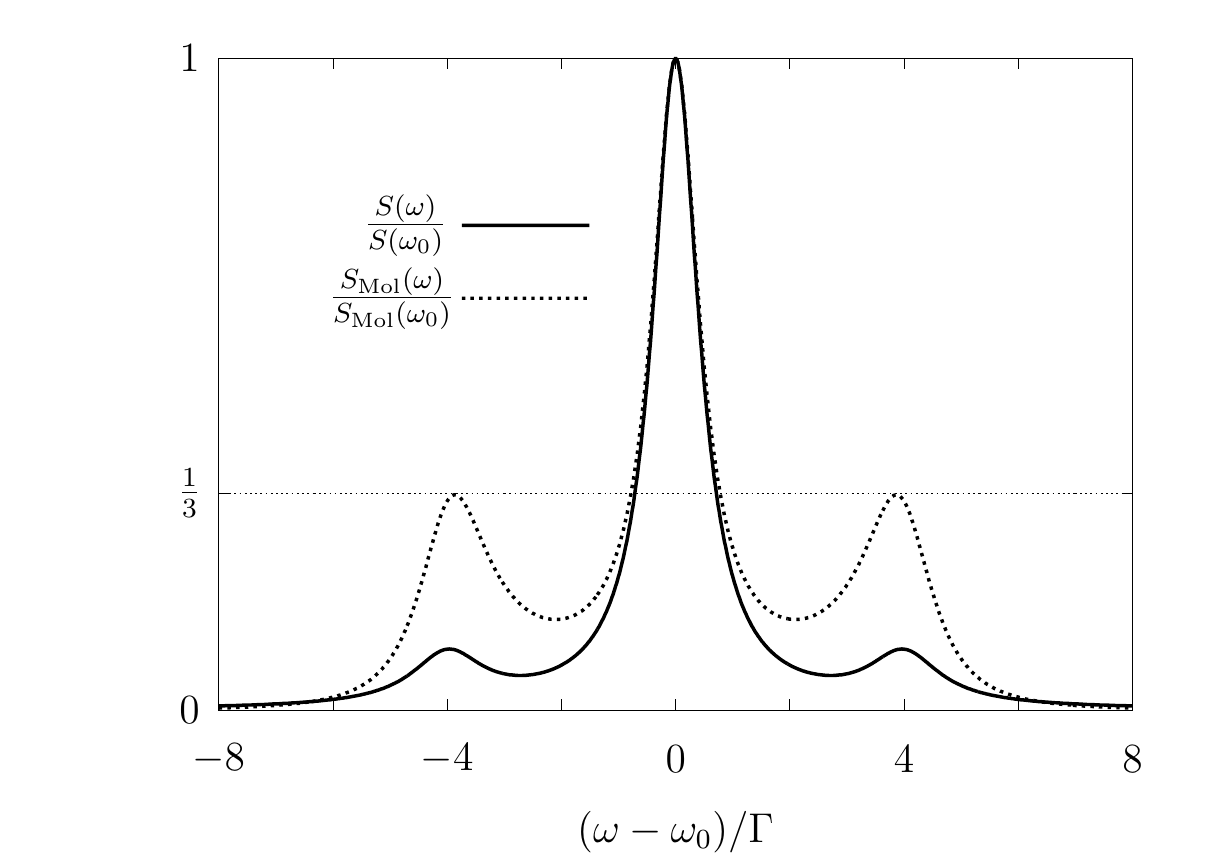}
\end{center}
\vspace*{-0.5cm}
\caption{The normalised power spectrum $S(\omega)/S(\omega_0)$ and the normalised Mollow triplet $S_{\rm Mol}(\omega)/S_{\rm Mol}(\omega_0)$ in Eqs.~(\ref{G6})--(\ref{p2}) as a function of $\omega $ for $\Gamma =10^8 \,$Hz, $\omega_0=10^{15}\,$Hz, and $\Omega=4 \, \Gamma$.} \label{spec}
\end{minipage}
\end{figure}

\begin{figure}[t]
\begin{minipage}{\columnwidth}
\begin{center}
\hspace*{-.75cm}
\includegraphics[scale=.8]{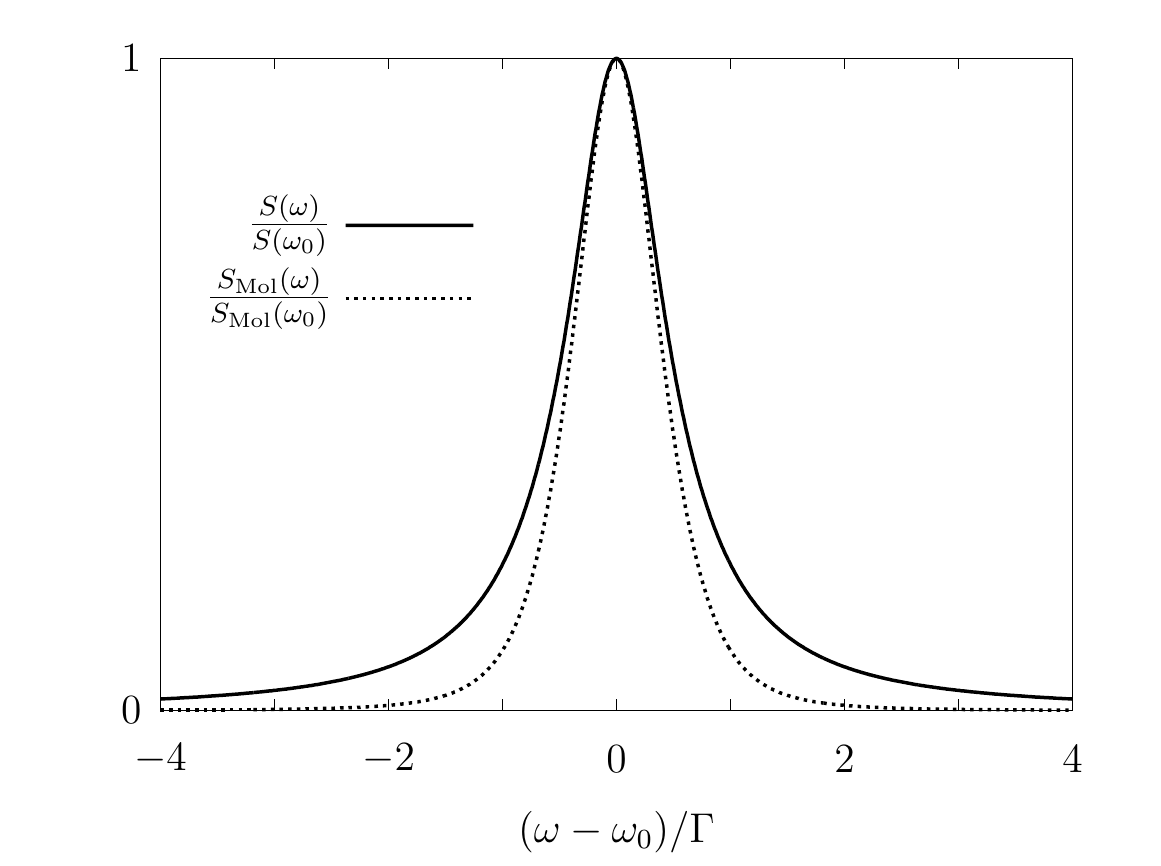}
\end{center}
\vspace*{-0.5cm}
\caption{The normalised power spectrum $S(\omega)/S(\omega_0)$ and the normalised Mollow triplet $S_{\rm Mol}(\omega)/S_{\rm Mol}(\omega_0)$ as a function of $\omega $ for the same $\Gamma$ and $\omega_0$ as in Fig.~\ref{spec} but for $\Omega=0.5 \, \Gamma$.} \label{spec3}
\end{minipage}
\end{figure}

Comparing the equations above with the equations in Section \ref{main}, we immediately see several similarities between the power spectrum $S(\omega)$ associated with the electric field and the Mollow spectrum $S_{\rm Mol}(\omega)$. Both spectra exhibit three peaks with the central peak located at $\omega = \omega_0$. In the case of sufficiently strong driving, two sidebands of equal height appear at $\omega = \omega_0 -\mu$ and $\omega = \omega_0 + \mu$. For weak driving, the sidebands vanish and there is only a single peak. The main difference is a significant reduction of the relative height of the sidebands and a different dependence of the overall amplitude on the laser Rabi frequency $\Omega$. This is illustrated in Figs.~\ref{spec} and \ref{spec3} which show the $\omega$-dependence of $S(\omega)$ (solid lines) and $S_{\rm Mol}(\omega)$ (dashed lines) for two different values of the laser Rabi frequency $\Omega$. In both figures the heights of the central peaks have been normalised to unity. When $\Omega$ is relatively small there is only a single central peak, but for sufficiently strong laser driving, both spectra have a three peak structure. This indicates that they both contain similar information about the atomic dynamics. However, the relative amplitudes of the sidebands of $S(\omega)$ are significantly smaller than the sidebands of $S_{\rm Mol}(\omega)$. 

\begin{figure}[t]
\begin{minipage}{\columnwidth}
\begin{center}
\hspace*{-.5cm}
\includegraphics[scale=.75]{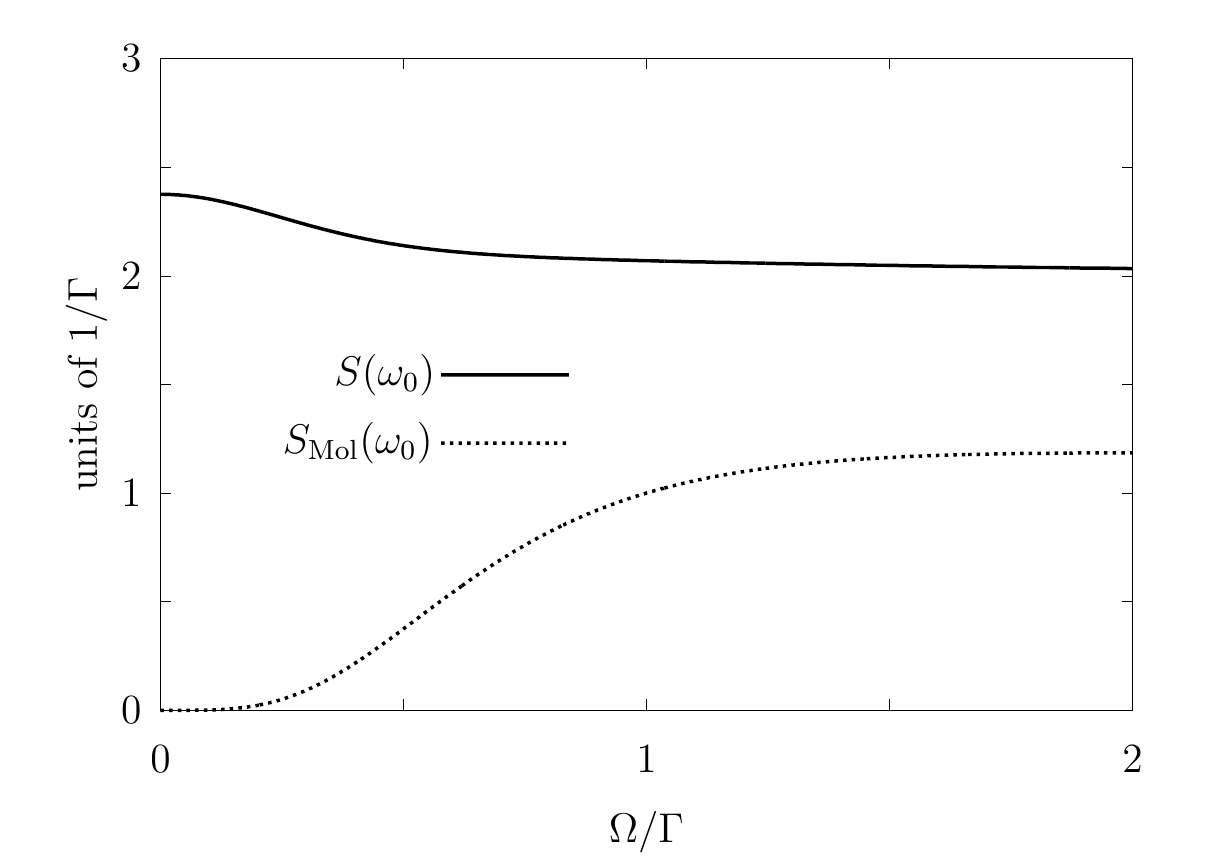}
\end{center}
\vspace*{-0.5cm}
\caption{Comparison of the unnormalised heights of the maximum peaks of $S(\omega)$ and $S_{\rm Mol}(\omega)$ for $\Gamma =10^8\,$Hz and $\omega_0=10^{15} \,$Hz.} \label{spec4}
\end{minipage}
\end{figure}

Another difference between $S(\omega)$ and $S_{\rm Mol} (\omega)$ is illustrated in Fig.~\ref{spec4}. This figure shows the dependence of the height of the central peaks of both spectra on the laser Rabi frequency $\Omega$. The Mollow triplet is usually normalised such that $\int {\rm d} \omega \, S_{\rm Mol} (\omega) $ is a direct measure of the stationary state photon emission rate $I_{\rm ss}$ of the laser-driven atomic system \cite{Breuer,Agarwal}. This means, its amplitude tends to zero, when $\Omega$ tends to zero. In contrast to this, $S(\omega)$ assumes its maximum when $\Omega$ tends to $0$. This means the spectrum $S(\omega)$ is not a measure of the photon emission intensity of the atomic source. A closer look at Eqs.~(\ref{xx}) and (\ref{yy}) shows that $\langle {\bf E}({\bf x}) \rangle_t$ is non-zero, even when the atomic system is in its ground state. When brought sufficiently close, the atomic dipole moment is expected to exert a force on a test charge. When this force is measured, the atomic state changes accordingly either into $|\lambda_0 (t) \rangle$ or into $|\lambda_1(t) \rangle$. We think that it would be interesting to observe this effect experimentally, in order to probe the dynamics of the laser-driven atom while assuming the presence of a photon-absorbing environment.

\subsection{Discussion}

The method usually used to derive Eq.~(\ref{GMol}) is quite different to the procedure we have used above in obtaining Eq.~(\ref{G5}). Within the usual approach one initially views the atom-field system as being {\em closed}. This entails solving the Heisenberg equations for the positive and negative frequency components of the electric displacement field in the multipolar gauge and EDA. One then invokes a rotating-wave approximation in order to elicit normal-ordering in the final expression in Eq.~(\ref{GMol}) \cite{Milonni}. Upon arriving at Eq.~(\ref{GMol}) for the Mollow spectrum one typically uses the Born-Markov master equation to calculate the expectation value in Eq.~(\ref{GMol}) itself.

If the rotating-wave approximation, which identifies the positive frequency component of the electric field with the lowering operator $\sigma^-$ is avoided, the correlation function $G_{\rm Mol}(\tau)$ is given by
\begin{align}\label{gmol3}
G_{\rm Mol}(\tau) &=\langle \sigma_{\rm x}(t+\tau)\sigma_{\rm x}(t)\rangle_{\rm ss}  \nonumber \\ & ={\rm Tr}[\sigma_{\rm x}(t+\tau){\cal T}_\tau( \sigma_{\rm x}(t)\rho_{\rm AI}(t))]\, .
\end{align}
Since the expectation value above is stationary it is independent of $t$, so time-averaging the right-hand-side of Eq. (\ref{gmol3}) leaves it invariant. Using Eq.~(\ref{sigspec}), $G_{\rm Mol}(\tau) $ in Eq.~(\ref{gmol3}) can therefore be written
\begin{align}\label{gmol2}
G_{\rm Mol}(\tau) =& \lim_{T \to \infty} {1 \over 2T} \, \int_{-T}^T {\rm d} t \, \sum_{i,j=0,1} \lambda_i \lambda_j \nonumber \\
& \times {\rm Tr} \left[ \, I \!\! P_j (t+\tau) \, {\cal T}_\tau \left( I \!\! P_i (t) \rho_{\rm AI}(t) \right) \, \right].
\end{align}
Apart from the absence of an additional projection operator $I \!\! P_i (t)$ within the argument of the superoperator ${\cal T}_\tau$, this expression is the same as the right-hand-side of Eq.~(\ref{G5}). The two expressions are equal if the density matrix $\rho_{\rm AI}(t)$ is diagonal in the $\{\ket{\lambda_i(t)}\}$ basis. This however, is not the case for the stationary state $\rho_{\rm ss}$ given in Eq.~(\ref{st}). Despite the similarity between Eqs.~(\ref{G5}) and (\ref{gmol2}), in general, one cannot write $G_{\rm Mol}(\tau)$ as a sum of conditional products of expectation values as in Eq.~(\ref{wsum}). Indeed, the operational meaning of an expression of the form $\langle f(t+\tau) f(t) \rangle$ is less forthcoming. The most obvious interpretation requires that within the experiment itself, the signal $f$ is physically split and one branch is delayed by a time $\tau$ before the branches are recombined.

In contrast to approaches that concentrate on calculating $G_{\rm Mol}(\tau)$, we have assumed that the entire atom-field system is open from the outset, and that it loses (free) energy due to the presence of the photon-absorbing environment. We have adopted this viewpoint at all levels of our calculation, in the sense that it is this viewpoint that leads us not only to Eq.~(\ref{G5}), but that also underlies our derivation of the Born-Markov master equation in Section \ref{meq} \cite{Ourpaper}. Furthermore, the correlation function that we calculate does not require the use of more elaborate experimental setups, which involve delaying part of the signal being measured.

\section{Conclusions} \label{conc}

This paper calculates the power spectrum $S(\omega)$ associated with the electric field ${\bf E} ({\bf x})$ generated by a laser-driven atomic two-level system. The detector at the position ${\bf x}$ should be placed a very small distance away from the atomic source at ${\bf 0}$. This is in principle feasible experimentally using currently available technology \cite{Usenko,Field,Petta,Elzerman,Poggio2,Clerk}. We have noted that the derived expression for $S(\omega)$ in Eq.~(\ref{fl1yx}) has some similarities with Mollow's resonance fluorescence spectrum \cite{cohen1,Breuer,Agarwal}, but that there are also several differences. For sufficiently strong laser driving there is a central peak as well as two sidebands. The relative height of the sidebands is significantly reduced in the case of the electric field spectrum (cf.~Figs.~\ref{spec} and \ref{spec3}). Moreover, the amplitude of this spectrum has a different dependence on the laser Rabi frequency $\Omega$ (cf.~Fig.~\ref{spec4}). It assumes its maximum when $\Omega$ tends to zero. 

The main result of this paper is to identify the power spectrum of the electric field generated by a laser-driven atomic system, while assuming the presence of a photon-absorbing environment. As a result of environment induced decoherence the radiation field which surrounds the atomic system is always in a mixed state of the vacuum and one-photon states, at least, over a coarse-grained time scale \cite{Hegerfeldt93,Molmer,Carmichael,Zoller}. As a result, the only non-zero contribution to the electric field signal $\langle {\bf E} ({\bf x}) \rangle_t$ comes from the atomic dipole moment ${\bf r}$. The predicted spectrum $S(\omega)$ contains a similar amount of information about the atomic system dynamics as Mollow's spectrum, but it can be calculated in a more direct way and constitutes an interesting alternative property. We hope that this paper provides new insight into the dynamics of spontaneously emitting quantum optical systems in the presence of decohering environments. The present discussion can be extended relatively easily to more complex systems and alternative measurement signals $f(t)$. \\[0.5cm]
\noindent {\em Acknowledgement.} This work was supported by the UK Engineering and Physical Sciences Research Council EPSRC.

\end{document}